\documentstyle[twocolumn,aps,prl,epsfig]{revtex}

\newcommand{\ybco}{$\mathrm{YBa_{2}Cu_{3}O_{7-\delta}}$}

\begin{document}

\draft
\author{A.V.Bondarenko, A.A.Prodan}
\address{
Karazin Kharkov National University, Physical Department, Svobody
sq. 4, 61077 Kharkov, Ukraine, E-mail:
Aleksander.V.Bondarenko@univer.kharkov.ua }
\author{Yu.T.Petrusenko, V.N.Borisenko}
\address{
National Science Center "Kharkov Institute of Physics and
Technology",Kharkov 61108, Ukraine}
\author{F.Dworschak, U.Dedek}
\address{
Institut fur Festk\"{o}rperforschung Forschungszentrum J\"{u}lich
GmbH, D-52425 J\"{u}lich, Germany}

\title{Effect of electron irradiation on vortex dynamics in \ybco\ single crystals}
\date{\today}

\twocolumn[\hsize\textwidth\columnwidth\hsize\csname@twocolumnfalse\endcsname

\maketitle

\begin{abstract}
We report on drastic change of vortex dynamics with increase of
quenched disorder: for rather weak disorder we found a single
vortex creep regime, which we attribute to a Bragg-glass phase,
while for enhanced disorder we found an increase of both the
depinning current and activation energy with magnetic field, which
we attribute to entangled vortex phase. We also found that
introduction of additional defects always increases the depinning
current, but it increases activation energy only for elastic
vortex creep, while it decreases activation energy for plastic
vortex creep.
\end{abstract}


]

The effect of random pinning on crystalline order and on dynamics
of the flux-line-lattice (FLL) was a subject of numerous
experimental and theoretical investigations. Neutron diffraction
~\cite{1} and $\mu$SR ~\cite{2} experiments gave experimental
evidence for the existence of two vortex solid phases with
different positional correlations. Magnetization measurements
~\cite{3} of $Bi_2Sr_2CaCu_2O_8$ showed that the field, $H_{on}$,
at which a steep increase of the measured current $J_m(H)$ begins,
coincides approximately with the field above which the intensity
of Bragg peaks sharply decreases ~\cite{1}. The field $H_{on}$ was
interpreted as a phase boundary between low field ordered and high
field disordered vortex phases. These experimental results are
supported by theoretical studies. It was shown that in presence of
rather weak disorder the FLL retains a quasi-long-range order
resulting in the so-called Bragg glass phase ~\cite{4}. However,
with the increase of random pinning or magnetic field a transition
to strongly disordered entangled vortex phase (glass phase) is
predicted ~\cite{5,6}.  A sharp increase in magnetization below
the fish-tail peak position $H_p$ was observed in
$Nd_{1.85}Ce_{0.15}CuO_{4-d}$ ~\cite{7} and non twinned \ybco\
~\cite{8} single crystals.

Magnetic measurements showed that in optimally doped \ybco\
crystals no fishtail behavior is observed both in detwinned
~\cite{a} and twinned ~\cite{9} samples, while decrease of the
oxygen content always induced non monotonous $J_m(H)$-curves. Two
distinctive peculiarities in the magnetization curves of oxygen
deficient ~\cite{9} and electron irradiated ~\cite{10} crystals
were observed: (1) the peaks $H_{on}$ and $H_p$ shift toward lower
magnetic fields with increasing defect concentration, and (2) in
magnetic fields $H < H_p$ the current $J_m$ increases, while  in
magnetic fields $H > H_p$ the current $J_m$ decreases with
increasing defect concentration. It is believed ~\cite{11,12} that
the peak $H_p$ separates elastic vortex creep in low and plastic
vortex creep, mediated by motion of the FLL dislocations, in high
magnetic fields. Thus the introduction of additional defects leads
to an increase of vortex pinning in the region of elastic creep
and such behavior is expected. On the other hand, decrease of
pinning force with increasing disorder observed in the region of
plastic creep is non trivial, and reasons of such behavior are not
known. The aim of this paper is to show the effect of point-like
defects concentration on vortex dynamics and pinning parameters in
\ybco\ single crystals.

The investigated sample was \ybco\ single crystal with $T_c$
$\approx$ 93.5 K and $\Delta T_c <$ 0.5 K. Twin planes (TP's)
inside the measured part of the sample were aligned in one
direction. The transport current was applied along the $ab$-plane
and at an angle $\alpha = 45^\circ$ with respect to TP's.
Measurements were performed in magnetic fields applied parallel to
the $c$-axis. Temperature stability of the sample during
measurements was better than 5 mK, and measurements in the normal
state showed that overheating of the sample at the highest
dissipation level of 50 $\mu$W did not exceed 10 mK.

Additional defects were introduced by irradiation with 2.5-MeV
electrons, which are suitable for production of point-like
defects. Irradiation was performed at temperatures $T\leq10 K$
~\cite{14} and after irradiation the current-voltage
characteristics (CVC) were measured without heating the sample
above 110 K. This excludes diffusion, and therefore annihilation
and clustering of the defects. Irradiation was performed at dose
rate $4.2\times10^{13} cm^{-2}sec^{-1}$ and at an angle $5^\circ$
off the c-axis to avoid electron channeling. Homogeneity of
electron beam was about 5\%. Following procedure used in
Ref.~\cite{b} we estimated that irradiation dose $10^{18} cm^{-2}$
produces the averaged over all sublattices concentration of the
defects $10^{-4}$ dpa, and that the penetration range $\sim$ 1mm
is at least two orders higher then thickness of the crystal $\cong
7\mu$m resulting in homogeneity of the defects along the c-axis
better then 1\%. Reduction of the $T_c$ after irradiation was
$\Delta T_{irr} =$ 0.65, 1 and 1.3 K after irradiation with doses
$10^{18}, 2\times 10^{18}$ and $3\times 10^{18} el/cm^2$,
respectively.

Fig.1 shows field variation of the measured "critical" current
$J_m$ determined at an electric field level $10^{-6} V/cm$. Before
irradiation the current $J_m$ continuously decreases with
increasing field as it was previously observed by Zhukov et al. in
crystals with very small oxygen deficiency, $\delta \simeq 0.03$
\cite{9}. After irradiation the $J_m(H)$ curves show fish-tail
behavior and the peak position $H_p$ in the $J_m(H)$ curves shifts
toward low fields with the dose. It is also seen that introduction
of additional defects increases $J_m$ for $H < H_p$, while the
current $J_m$ decreases with the dose for $H > H_p$. Such
variation of the $J_m$ and $H_p$ with increasing of the defect
concentration is analogous to the behavior previously observed in
oxygen deficient \cite{9} and in electron irradiated \cite{10}
\ybco\ single crystals.
\begin{figure}
\epsfig{file=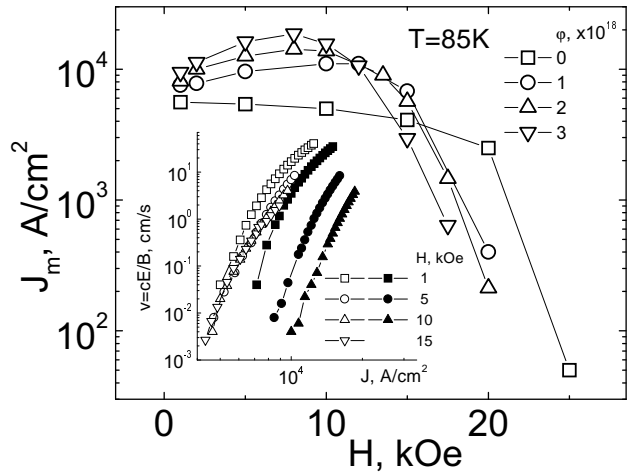,clip=,silent=,width=3.3in} \caption{Field
variation of the current $J_m$. The inset shows vortex velocity
versus $J$ before (open symbols) and after (dark symbols)
irradiation with $\varphi = 10^{18}$ el/cm$^2$.} \label{fig1}
\end{figure}
In low magnetic fields ($H \leq 15 kOe$ for non irradiated sample,
and $H < H_p$ for irradiated samples) the CVC data follow the
dependence ~\cite{e}
\begin{equation}
E = E_0exp[-(U_0/kT)(J_c/J)^\mu], \label{eq1}
\end{equation}
where the exponent $\mu \cong 1$, $E_0$ is a constant, $U_0$ is
the activation energy, and $k$ is the Boltzmann constant. The
depinning critical current $J_c$ can be determined by
extrapolation of the ratio $\rho_d(J)/\rho_{BS}$ to unity
\cite{12} assuming that at current density $J = J_c$ the
differential resistivity $\rho_d \equiv dE/dJ$ equals the flux
flow resistivity in the Bardeen-Stephen model, $\rho_{BS} =
\rho_NB/B_{c2}$ \cite{15}, where $\rho_N$ is the normal state
resistivity. Induction of the upper critical field was estimated
assuming that $B_{c2} = (dB_{c2}/dT)(T - T_c)$ with $dB_{c2}/dT$ =
-2.5T/K \cite{c}. Field variation of the $J_c$ is shown in the
inset of Fig.\ref{fig2}a. Substituting these values of $J_c$ in
Eq.\ref{eq1}, and fitting experimental $E(J)$-curves by this
equation we derived $U_0(H)$ dependence shown in the inset of
Fig.\ref{fig2}a.
\begin{figure}
\epsfig{file=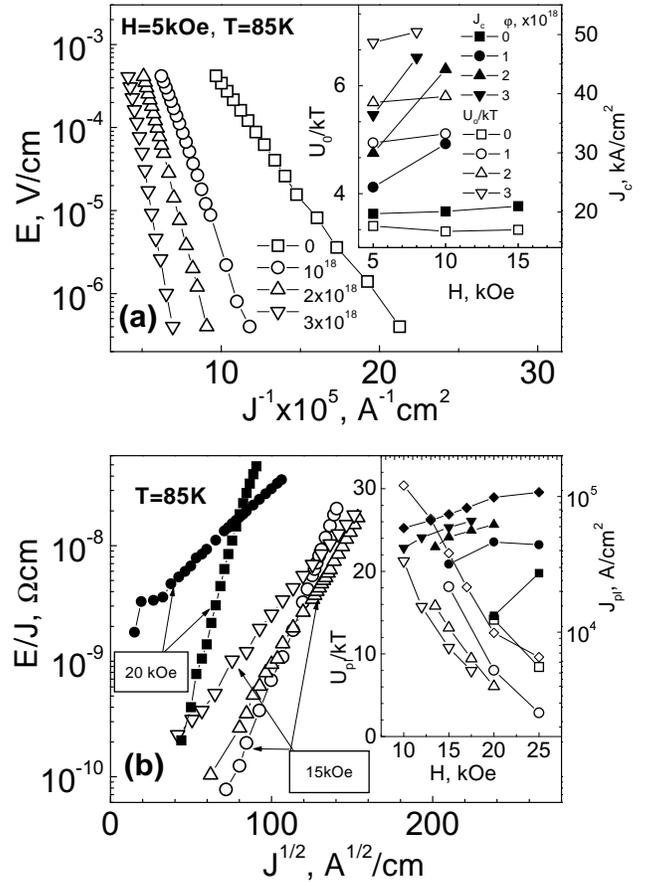,clip=,silent=,width=3.3in} \caption{(a) The
$E(J)$ curves plotted as $E(J) vs. 1/J$. The inset shows field
variation of the $J_c$ and $U_0$. (b) The $E(J)$ curves measured
before (squares) and after irradiation with doses $10^{18}$
(circles), $2\times10^{18}$ (up triangulars), and $3\times10^{18}
el/cm^2$ (down triangulars), and plotted as $E(J)/J vs. J^{1/2}$.
The inset shows $J_{pl}(H)$-dependence (dark symbols) and
$U_{pl}(H)$-dependence (open symbols) derived for $T = 85 K$
before irradiation (squares) and after irradiation with doses
$10^{18}$ (circles), $2\times10^{18}$ (up triangulars), and
$3\times10^{18} el/cm^2$ (down triangulars), and derived for $T =
82 K$ and $\varphi = 3\times10^{18} el/cm^2$ (diamonds).}
\label{fig2}
\end{figure}
As one can see in Fig.\ref{fig1} and in the inset of
Fig.\ref{fig2}a, vortex dynamics, namely field variation of the
pinning parameters, strongly depends on the strength of disorder.
In presence of weak disorder (before irradiation) $J_c$, $U_0$,
and vortex velocity $v = cE/B$ do not depend on magnetic field
indicating a single vortex creep regime, as it is predicted by the
collective pinning theory \cite{e} for low magnetic fields. Also,
for this regime of vortex creep the current $J_m$ decreases with
increasing magnetic field due to increase of the flux density.
Increase of the disorder induces field variation of the pinning
parameters. After irradiation $J_c$ and $U_0$ increase with
magnetic field that leads to reduction of vortex velocity and to
increase of the current $J_m$ with increasing magnetic field.
These observations indicate that before and after irradiation we
test different vortex phases.

In presence of weak quenched disorder the ordered Bragg-glass
phase is expected \cite{4} and our experimental data probably
indicates that dynamics of just this vortex phase is described by
the single vortex creep regime. Also note, that derived exponent
$\mu \cong 1 \pm 0.15$ is close to the value of $\mu \cong 0.7 -
0.8$ predicted for the Bragg-glass phase for not too small
currents \cite{6}. With increase of disorder a transition of the
Bragg-glass to entangled vortex phase is expected \cite{5,6}. The
derived after irradiation weak increase of the $U_0$ and rapid
increase of the $J_m$ with magnetic field correlates with previous
experimental findings \cite{8,10} for magnetic fields $B >
B_{on}$, or in magnetic fields where the entangled vortex solid is
expected. Therefore we believe that after irradiation we test the
entangled vortex phase and dynamics of this phase is characterized
by the increase of both $J_c$ and $U_0$ with increasing magnetic
field, which lead to rapid increase of the current $J_m$. Also,
introduction of additional defects increases both the $U_0$ and
$J_c$ resulting in steep increase of the current $J_m$ with
irradiation dose, see Fig.\ref{fig1}, in agreement with previous
investigations \cite{10,b}.

\begin{figure}
\epsfig{file=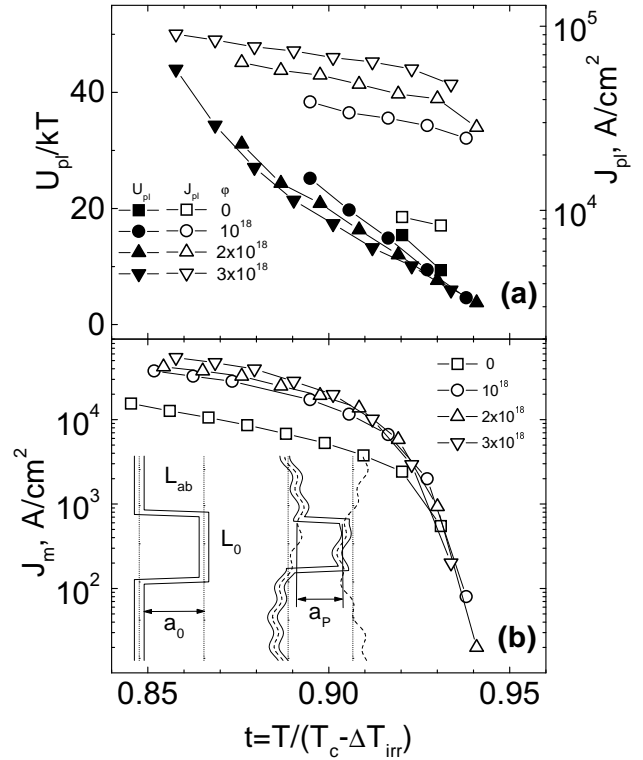,clip=,silent=,width=3.3in} \caption{(a)
Temperature variation of $U_{pl}$ and $J_{pl}$ for $H = 15 kOe$.
(b) Temperature variation of $J_m$ for $H = 15 kOe$. The inset
shows nuclei for the plastic creep in the absence (left) and in
the presence (right) of random pinning.} \label{fig3}
\end{figure}

Let us consider experimental data obtained in magnetic fields $H
\geq 20 kOe$ for nonirradiated sample, and in magnetic fields $H >
H_p$ for irradeated samples. As one can see in Fig.\ref{fig2}b,
the CVC data follow the dependence predicted for motion of the FLL
dislocations \cite{11,17}
\begin{equation}
E(J) = \rho_0Jexp\{-(U_{pl}/kT)[1-(J_{pl}/J)^\mu]\}, \label{eq2}
\end{equation}
where $\mu$ = -1/2, and $\rho_0$ is a constant. The main
peculiarities of the curves presented in Fig.\ref{fig2}b are that
the slope of the curves measured in the same magnetic field
decreases with increasing  defects concentration and that these
curves intersect one another. For vortex creep described  by
Eq.\ref{eq2} such behavior is possible only in that case when the
critical current increases, but the activation energy decreases
with increasing defect concentration. Extrapolating the ratio
$\rho_d(J)/\rho_{BS}$ to unity we derived field and temperature
variation of the current $J_{pl}$ shown in the inset of
Fig.\ref{fig2}b and in Fig.\ref{fig3}a, respectively. Substituting
these values of $J_{pl}$ in Eq.\ref{eq2}, and fitting experimental
$E(J)$-curves by this equation we determined the field and
temperature variation of $U_{pl}$ shown in the inset of
Fig.\ref{fig2}b and in Fig.\ref{fig3}a, respectively.

The current $J_{pl}$ is determined by the interaction of the
dislocations with the FLL and random pinning centers. Interaction
with the FLL results in a shear limited contribution \cite{18}
$J_{sh} \propto c_{66}/Bd \propto B^{1/2}$, where $c_{66} =
\Phi_0B/(8\pi\lambda)^2$ is the shear modulus, $\Phi_0$ is the
flux quantum, $\lambda$ is the penetration depth, $d \approx a_0$
is the width of channel for moving dislocation, and $a_0 \approx
(\Phi_0/B)^{1/2}$ is the intervortex distance. As one can see in
the inset of Fig.\ref{fig2}b, in magnetic fields not very close to
the melting point the current $J_{pl}$ really increases with the
field in agreement with theoretical predictions. In the presence
of random pinning the shear modulus $c_{66}$ decreases \cite{13}
and therefore $J_{sh}$ also decreases. However, the contribution
of the core pinning increases with the defect concentration
\cite{e}. The derived increase of the current $J_{pl}$ with
irradiation dose indicates that increase of the core pinning
dominates over reduction of the $J_{sh}$.
The activation energy $U_{pl}$ decreases with the increase of both
irradiation dose and magnetic field. A  decrease of the activation
energy with increasing field agrees with theoretical calculations,
$U_{pl} \approx \varepsilon\varepsilon_0a_0 \propto B^{-1/2}$
\cite{19}, where $\varepsilon_0 = (\Phi_0/8\pi\lambda)^2$, and
previous experimental findings \cite{11,12}. But, a decrease of
the activation energy with increasing irradiation dose is not
evident. Of course, the $T_c$ decreases after irradiation which
leads to an increase of the $\lambda(T) =
\lambda(0)/[1-(T/T_c)^2]^{1/2}$, and hence to a decrease of the
activation energy $U_{pl} \propto \lambda^{-2}$. However, the
reduction of the $T_c$ can not describe the fast decrease of
$U_{pl}$ in our measurements. This is demonstrated in
Fig.\ref{fig3}a, which shows dependence $U_{pl}$ vs. $t \equiv
T/(T_c - \Delta T_{irr})$. It is evident that $U_0$ decreases with
the dose increasing even if we take into account the reduction of
$T_c$.

To explain this behavior let us consider a displacement of a
vortex segment $L_0$ over intervortex distance $a_0$ as shown in
the left-hand inset of Fig.\ref{fig3}b. In the absence of random
pinning the energy of such nucleus can be written as $U_{pl} =
2E_{el} + E_{pl}$, where $E_{el} \approx
\varepsilon\varepsilon_0a_0$ is elastic energy of the vortex
segment $L_{ab}$, and $E_{pl}$ is the energy required for
displacement of the vortex segment $L_0$ over the distance $a_0$.
At small driving forces such a nucleus is stable when $E_{pl} \geq
2E_{el}$. Thus we obtain a minimal activation energy $U_{pl}
\approx 4\varepsilon\varepsilon_0a_0$, which coincides within a
factor 4 with previous estimates $U_{pl} \approx
\varepsilon\varepsilon_0a_0$ \cite{19}. In the absence of random
pinning the equilibrium positions of the vortices are the straight
lines as it is shown by two parallel straight lines. In the
presence of random pinning the equilibrium positions become curved
as shown by dashed lines in the right-hand inset of
Fig.\ref{fig3}b. Therefore, along the vortex lines a vortex
fragments, for which an average distance $a_p$ between two
neighboring equilibrium positions of vortex lines is smaller then
$a_0$ appear. In this case the energy $U_{pl} \approx
4\varepsilon\varepsilon_0a_p$ is smaller then activation energy in
the absence of pinning because $a_p < a_0$. Substantial
displacements of vortex segments, $\Delta a_0 = a_0 - a_p \simeq
0.2a_0 \simeq 8 nm$, due to core interaction with individual
pinning centers probably unreliable, and we attribute them to
fluctuations of the defects concentration, which naturally present
in real crystals. Indeed, equating the work $A =
J_{pl}F_0L_0a_p/c$, required for displacement of the segment $L_0$
over distance $a_p$, to the elastic energy, $2E_{el} = U_{pl}/2$,
we estimated $L_0 \cong$ 500 nm for $\varphi = 10^{18}$ el/cm$^2$,
and derived that for coherence length $\xi$(85K) = 5 nm the
segment $L_0$ interacts with about 200 point defects. Therefore,
fluctuations of the defects concentration of 10\% can give
difference in core interaction with about 20 point defects in
different positions of the segment $L_0$.

TP's strongly affect pinning and dynamics of vortices
~\cite{f,21,22}. In particular, being plane defects they form
channels of easy vortex motion along the plane of twins
~\cite{21,22}. Due to suppression of superconducting order
parameter within the TP's some part of vortices is trapped by the
TP's ~\cite{20}, and pinning of these vortices along the TP's may
be reduced as compared with pinning in the bulk of the crystal
~\cite{21}. Therefore, for the same driving force velocity of the
trapped vortices can be higher compared with velocity of vortices
placed in the bulk of crystal. In high magnetic fields
contribution of the trapped vortices to dissipation of energy is
small due to small fraction of these vortices. However, in a
magnetic field 1 kOe the intervortex separation $a_0 \cong 140 nm$
becomes comparable with the distance between twins $d \cong 300
nm$ in our sample, and a significant part of vortices can be
trapped by the TP's. Therefore, contribution of the trapped
vortices increases with decreasing magnetic field, and deviation
from the field scaling for non irradiated sample in low fields
presented in the inset of Fig.\ref{fig1} probably reflects the
reduced pinning of the trapped vortices.

In conclusion, we have shown that vortex dynamics dramatically
depends on the strength of disorder. In presence of weak disorder
a single vortex creep is realized, which we attribute to dynamical
property of the ordered Bragg-glass phase. In presence of strong
disorder and for elastic vortex creep we have found rapid increase
of the depinning current and weak increase of the activation
energy with increasing magnetic field, which we attribute to
dynamical property of the entangled vortex solid. We also found
that the introduction of additional defects always increases the
depinning critical current, but it increases the activation energy
only for the elastic vortex creep, while it decreases the
activation energy for the plastic vortex creep.

We gratefully acknowledge support by German Federal Ministry for
Research and Technology under the Project UKR-032-96.

%
%

\end{document}